\documentclass[11pt]{article}
\usepackage{vietnam,epsfig}

\def\Journal#1#2#3#4{{#1} {\bf #2}, #3 (#4)}

\def\PLB{{\em Phys. Lett.}  B}
\def\PRL{\em Phys. Rev. Lett.}
\def\PRD{{\em Phys. Rev.} D}

\def\EPJC{{\em Eur. Phys. J.} C}

\def\be{\begin{equation}}
\def\ee{\end{equation}}
\def\bea{\begin{eqnarray}}
\def\eea{\end{eqnarray}}

\begin{document}
\vbox{\hfill {CLNS 04/1902}}
\vspace*{3.5cm}
\title{CKM RESULTS FROM CLEO}

\author{ Karl M.~Ecklund }

\address{F.~R.~Newman Laboratory for Elementary-Particle Physics,\\
Cornell University, Ithaca, NY 14850, USA}

\maketitle\abstracts{
I present recent CLEO measurements of inclusive semileptonic $B$ decay
spectra and the branching fraction.  I also report preliminary results
from a combined fit to spectral moments, which are compared to
expectations in the Heavy Quark Expansion and used to precisely
determine $|V_{cb}|$.} 

\section{Introduction}
Measurements of the CKM matrix elements $V_{ub}$ and $V_{cb}$ provide
important constraints on the weak flavor sector of the Standard Model.
In the familiar Unitarity Triangle, which displays a test of the
unitarity of the CKM matrix, the ratio $|V_{ub}|/|V_{cb}|$ defines a
annulus centered at $\rho=0,\eta=0$, complementary to information,
e.g., from $B^0-\bar B^0$ mixing and the CP asymmetry in $B\to J/\psi
K^0$ decays.  The magnitude of these CKM matrix elements may be
measured from the rate of semileptonic $B$ decays.  At present the
precision is limited by hadronic effects.  Within the framework of the
Heavy Quark Expansion (HQE), the hadronic uncertainties may be
constrained using the moments of inclusive decay spectra.  The spectra
also allow a test of the success of the HQE.  Below I present recent
CLEO measurements of inclusive semileptonic $B$ decay and the status
of $|V_{cb}|$ extraction using the HQE.

\section{Invariant Mass Spectra}\label{sec:MX}
\begin{figure}
\begin{center}
\epsfig{figure=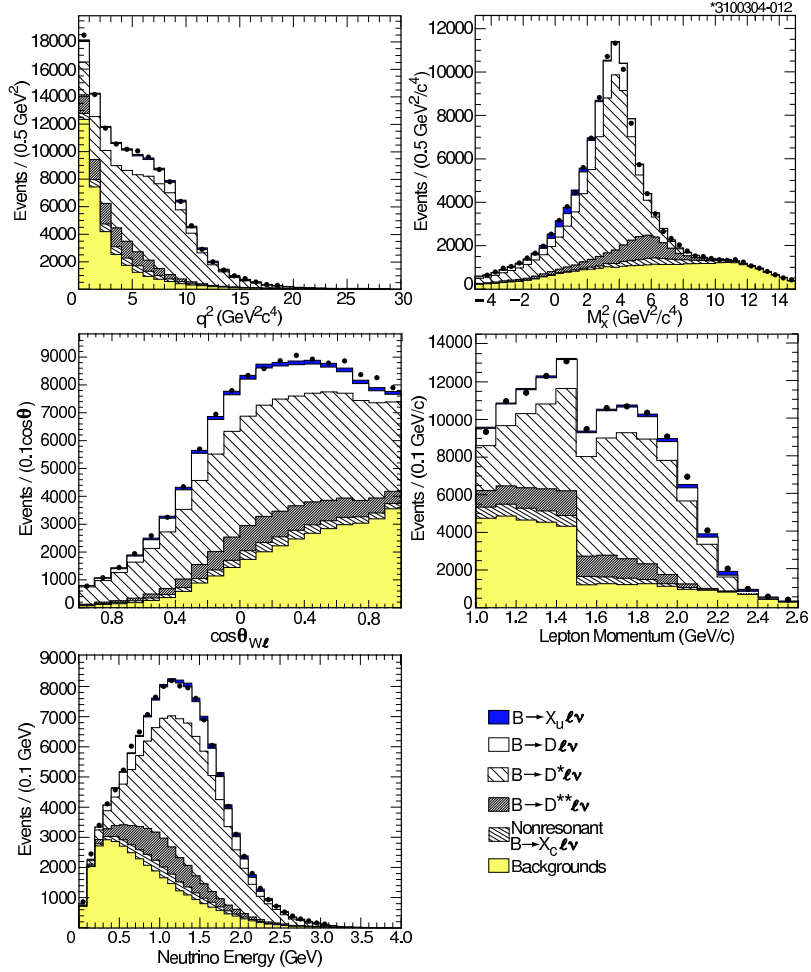,width=0.67\textwidth}
\end{center}
\caption{Projections of fit to the triple differential distribution
${d^3\Gamma \over dq^2 dM^2_X d\cos\theta_{W\ell} }$.
\label{fig:TripleDifferential}}
\end{figure}

From $\Upsilon(4S)\to B\bar B$ data, CLEO recently measured moments of
the hadronic ($M_X^2$) and leptonic ($q^2$) invariant mass-squared
spectra in $B\to X_c\ell\nu$ using a neutrino reconstruction
technique,~\cite{MX} in which the four momentum of a neutrino candidate is
inferred from missing energy and momentum, relying on the hermeticity of
the CLEO detector.  Using only the 
neutrino and charged lepton ($e$,$\mu$) four momenta, $q^2$ is
the square of the lepton-neutrino invariant mass and the hadronic
recoil mass is approximated by ignoring the final term in
Equation~\ref{eq:MX2}, which is small because the $B$ is nearly at rest.
\begin{eqnarray}
M_X^2 & = & M_B^2 + q^2 -2 E_{\rm beam} (E_\ell + E_\nu) + 
2 |{\mathbf p_B}||{\mathbf{p}_\ell+\mathbf{p}_\nu}| cos \theta_{B-\ell\nu}
\nonumber \\
& \approx & M_B^2 + q^2 -2 E_{\rm beam} (E_\ell + E_\nu)
\label{eq:MX2}
\end{eqnarray}
We then make a maximum likelihood fit to the triple differential decay
distribution.  Projections of the fit are shown in
Figure~\ref{fig:TripleDifferential}.  While the fit-determined
branching fractions are model dependent, moments of the inclusive
spectra are much less so.  We extract first and second moments of
$M_X^2$ and $q^2$, subject to two cuts on the charged lepton energy:
$E_\ell> 1.0,1.5$ GeV.

The results (Table \ref{tab:hqe}) are consistent with an earlier CLEO
measurement~\cite{cleohad} and comparable to a recent {\sc BaBar}
measurement.\cite{babar.mx} 

\section{Lepton Energy Spectrum and Branching Fraction}\label{sec:El}
\begin{figure}
\begin{center}
\epsfig{figure=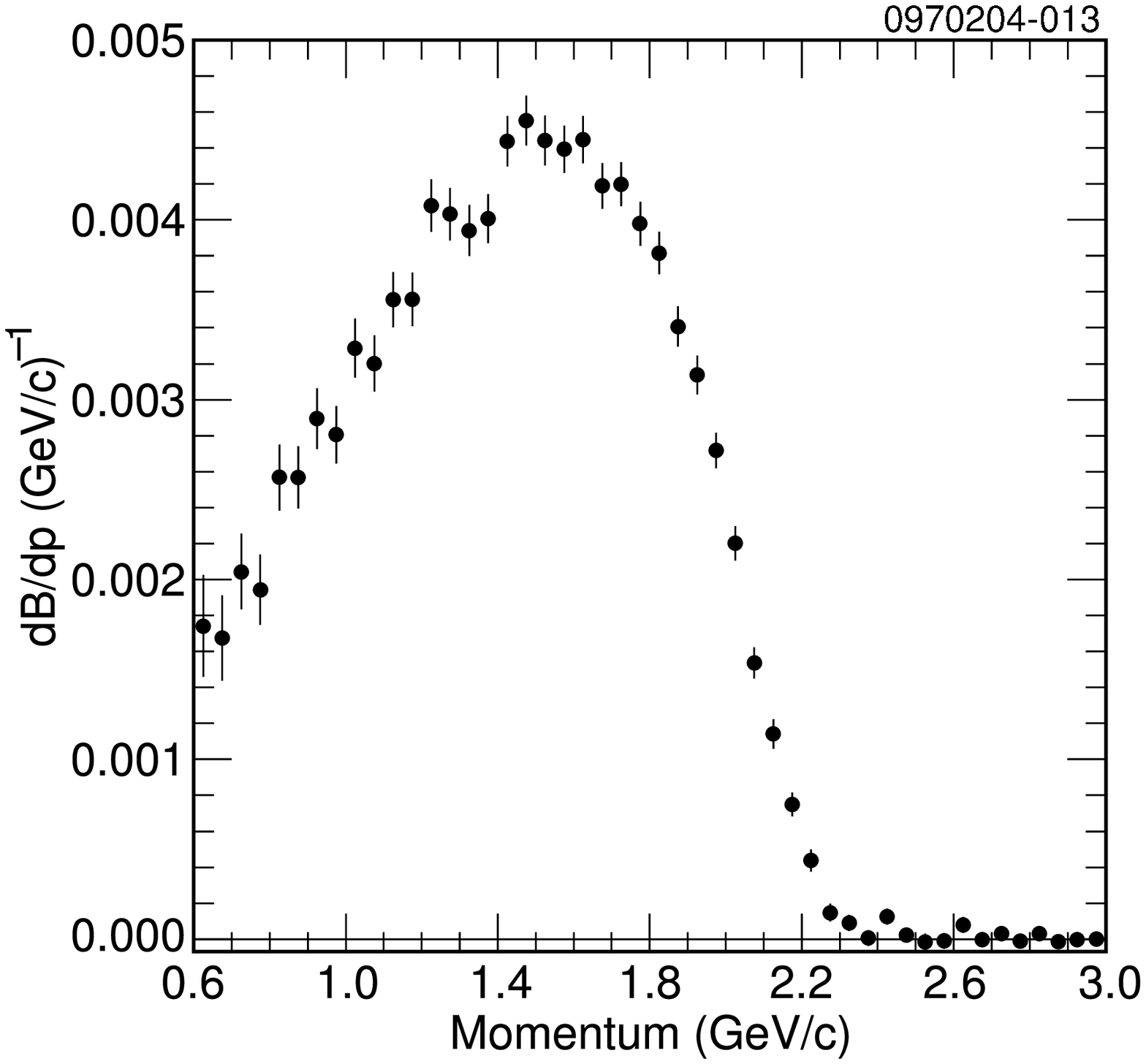,width=0.45\textwidth}\hfill
\epsfig{figure=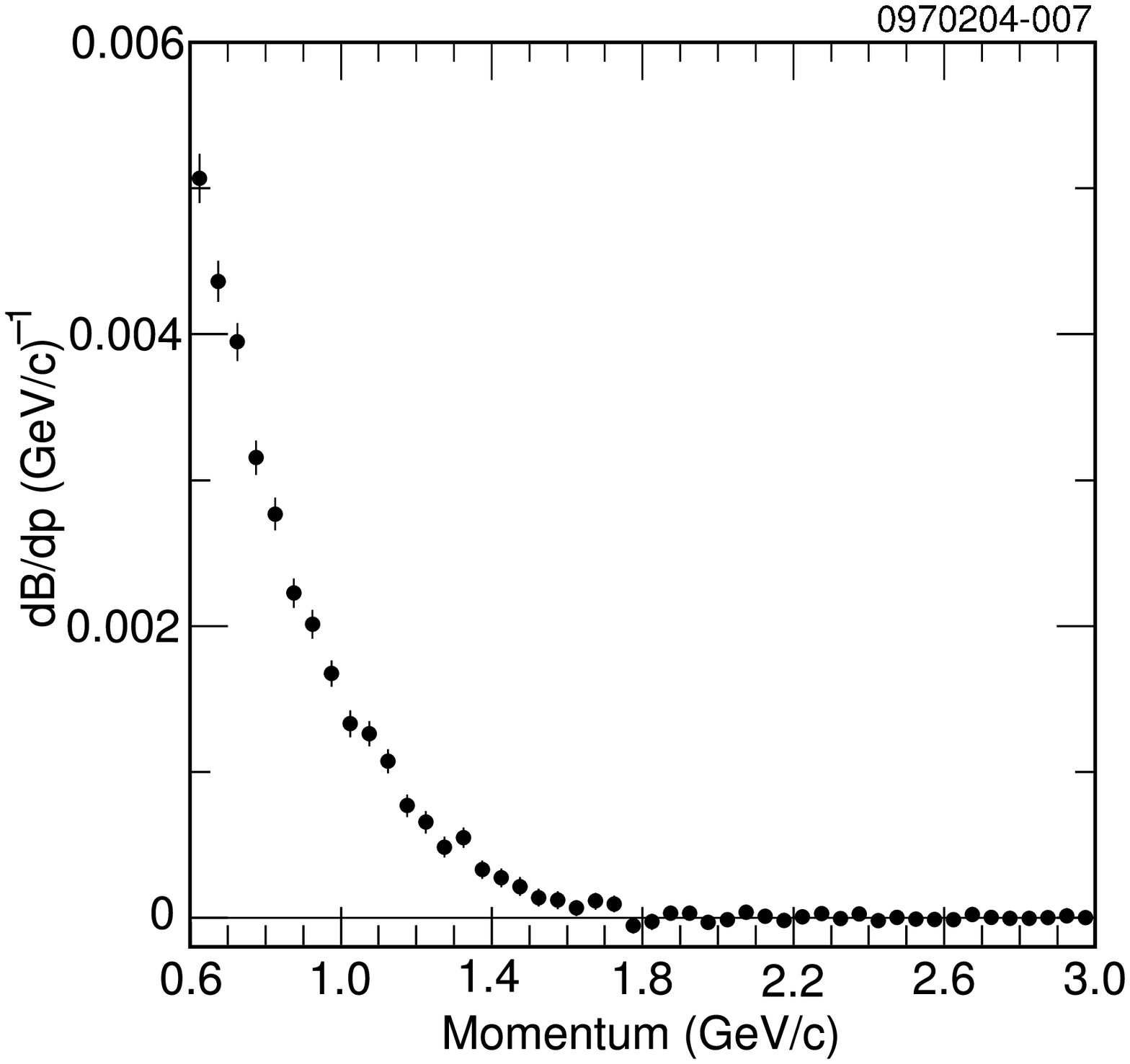,width=0.45\textwidth}
\caption{$B\to X\ell\nu$ lepton spectrum (left) and charm secondary lepton
         spectrum (right). \label{fig:LeptonEnergy}}
\end{center}

\end{figure}

In a second analysis~\cite{El} CLEO uses a dilepton sample to measure
the inclusive semileptonic branching fraction and electron spectrum
for $p>0.6$ GeV/$c$.  A high momentum ($p>1.4$ GeV$/c$) $\mu$ or $e$
candidate isolates a 97\% pure sample of $\Upsilon(4S)\to B\bar B$
decays.  The charge of this lepton tag identifies the flavor of one
$B$ in the event.  The spectrum of additional electrons has
contributions from semileptonic decay of the other $B$ (signal),
semileptonic decays of charm from the either of the $B$'s daughters,
and other small backgrounds from, e.g., photon conversions and
$J/\psi\to ee$.  Using charge and kinematic correlations, we separate
primary ($b\to c e^-\bar\nu$) from secondary ($b\to c$, $c\to q
e^+\nu$) electrons. Electrons from the decay of charm daughters of the
same $B$ that produced the tag are efficiently vetoed with cuts on
charge and direction relative to the tagging lepton.  In $B^+B^-$ events,
like-sign leptons identify secondary backgrounds.  For $B^0 \bar B^0$
events, mixing will also lead to like-sign events.  The effects of
mixing can be unfolded statistically, leading to efficiency-corrected
measurements of the primary and secondary electron momentum spectra
(Fig.~\ref{fig:LeptonEnergy}).  By integrating the primary spectrum
above 0.6 GeV$/c$, and making a small extrapolation below, we measure
an inclusive branching fraction of $(10.91 \pm 0.09 \pm 0.24)$\%.
For later use we also report first and second moments of the $B\to X_c
e\nu$ spectrum with three cuts $E_\ell>0.7,1.2,1.5$ GeV (Table~\ref{tab:hqe}),
obtained by subtracting the small $B\to X_u e\nu$ component. 

\section{Global fit for $|V_{cb}|$}
\begin{table}
\caption{Input moment measurements for $B\to X_c\ell\nu$ and $B\to
X_s\gamma$ and resulting HQE parameters from combined fit.  The
subscripts indicate lepton or photon energy cuts.  All units are
powers of GeV.}
\label{tab:hqe}
\begin{center}\begin{tabular}{|rl|rl|rl|}
\hline
\multicolumn{2}{|c|}{Input First Moments} & 
\multicolumn{2}{|c|}{Input Second Moments} &
\multicolumn{2}{|c|}{Output} \\ \hline
$\left< q^2 \right>_{1.0}$ & $4.89 \pm 0.14$ & & & & \\
$\left< q^2 \right>_{1.5}$ & $5.29 \pm 0.12$ & & & 
$m_b$  & $4.564^{+0.073}_{-0.074}$ \\ 

$\left<M_X^2-M_{\bar D}^2\right>_{1.0}$ & $0.46 \pm 0.12 $ &
$\left<\left(M_X^2-\left<M_X^2\right>\right)^2\right>_{1.0}$ 
	& $1.27 \pm 0.45 $ &
$m_c$ & $1.16^{+0.10}_{-0.11}$ \\ 

$\left<M_X^2-M_{\bar D}^2\right>_{1.5}$ & $0.29 \pm 0.07 $ &
$\left<\left(M_X^2-\left<M_X^2\right>\right)^2\right>_{1.5}$ 
	& $0.63 \pm 0.15 $ &
$\mu_\pi^2$ & $0.483^{+0.050}_{-0.054}$ \\ 

$\left<E_\ell\right>_{0.7}$ & $1.451 \pm 0.009$ &
$\left<(E_\ell-\left<E_\ell\right>)^2\right>_{0.7}$ & $0.137\pm 0.002$ & 
$\mu_G^2$ & $0.105^{+0.083}_{-0.081}$ \\ 

$\left<E_\ell\right>_{1.2}$ & $1.631 \pm 0.004$ &
$\left<(E_\ell-\left<E_\ell\right>)^2\right>_{1.2}$ & $0.064\pm 0.002$ & 
$\tilde\rho_D^3$  & $0.91 \pm 0.03$ \\ 

$\left<E_\ell\right>_{1.5}$ & $1.779 \pm 0.004$ &
$\left<(E_\ell-\left<E_\ell\right>)^2\right>_{1.5}$ & $0.032\pm 0.001$ & 
$\rho_{LS}^3$  & $0.05 \pm 0.17$ \\ 

$\left<E_\gamma\right>_{2.0}$ & $2.346 \pm 0.034$ & 
$\left<E_\gamma^2-\left<E_\gamma\right>^2\right>_{2.0}$ & $0.023 \pm 0.007$ &
& \\

\hline
\end{tabular}

\end{center}
\end{table}

\begin{figure}
\epsfig{file=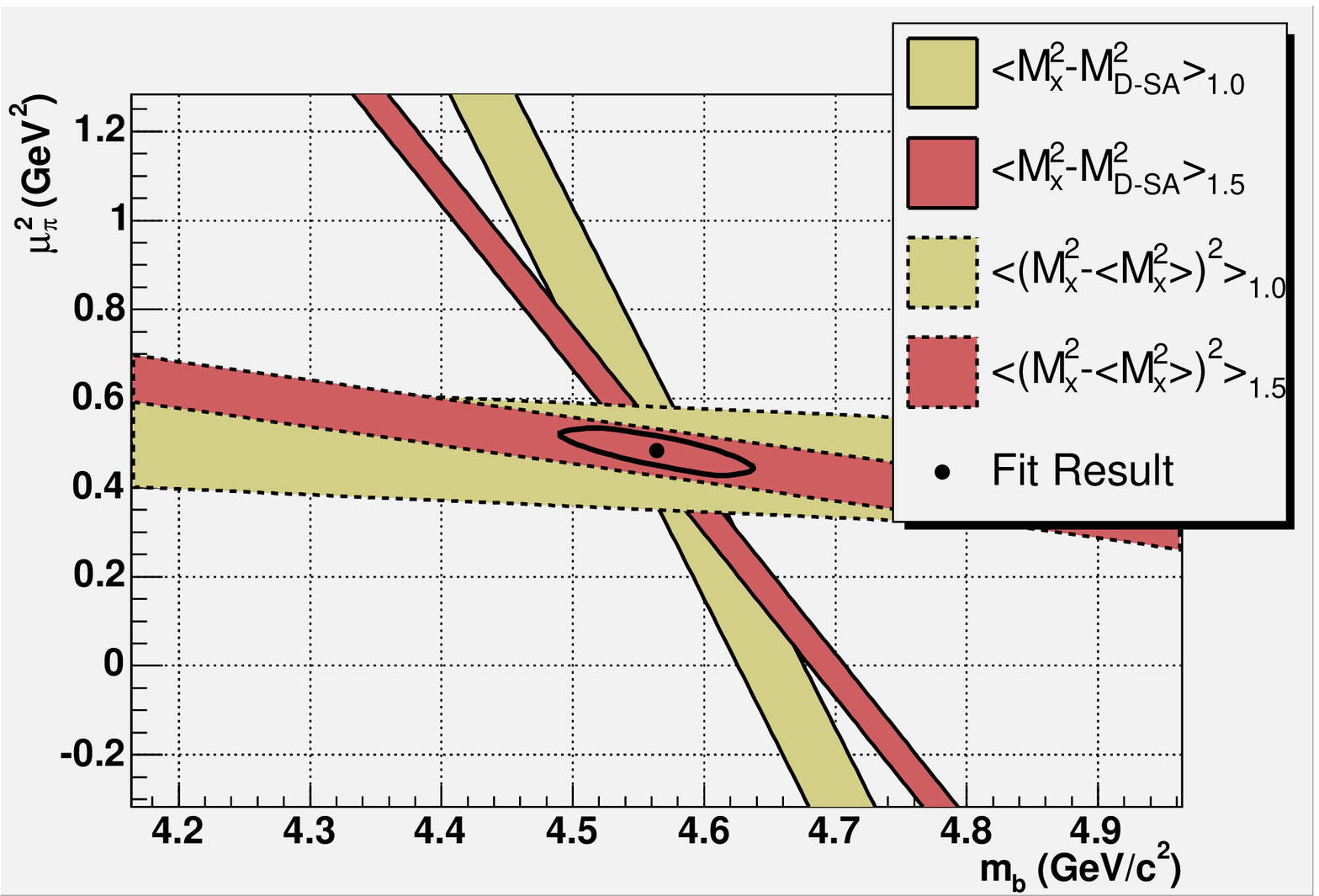,width=0.495\textwidth} \hfill
\epsfig{file=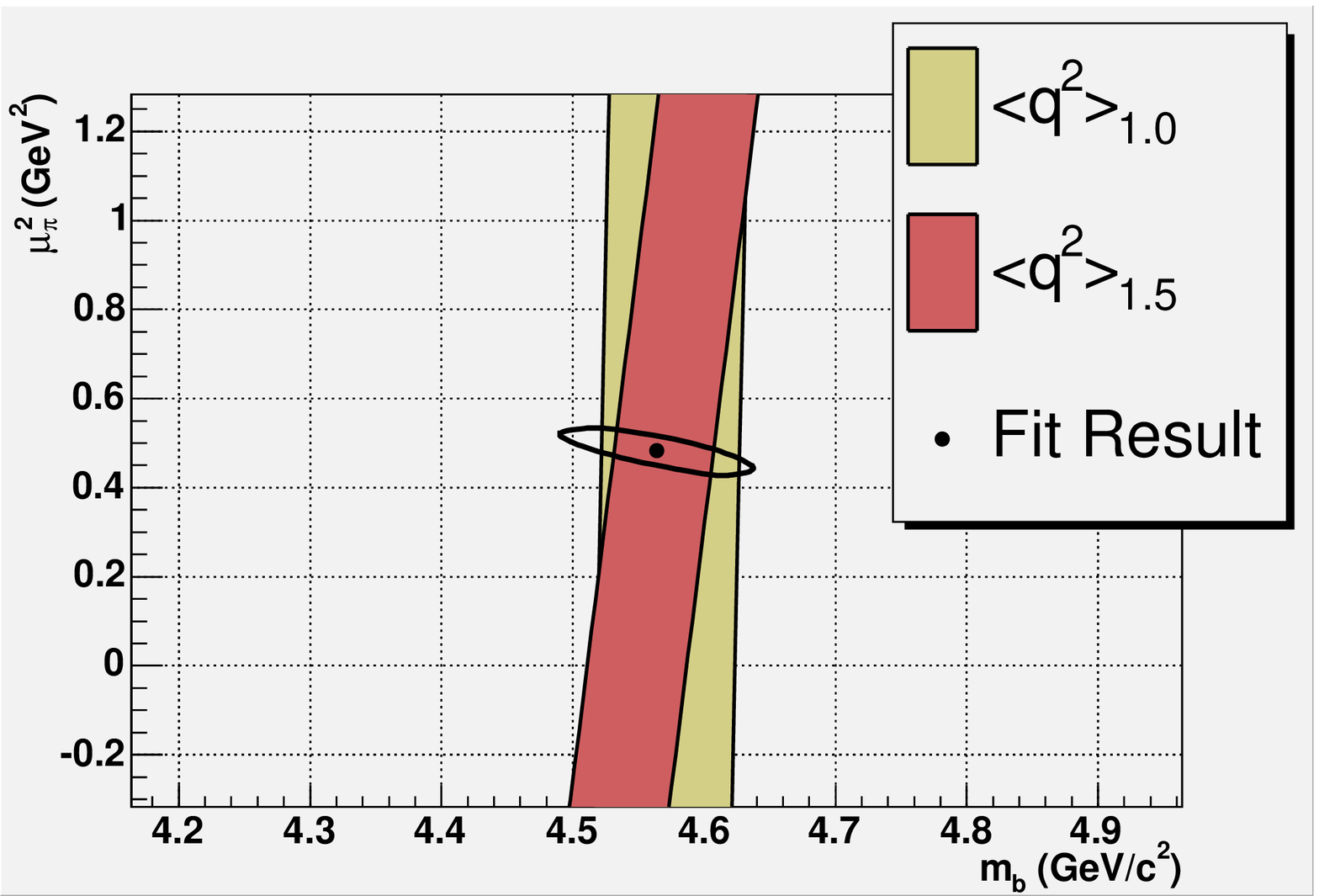,width=0.495\textwidth} \\
\epsfig{file=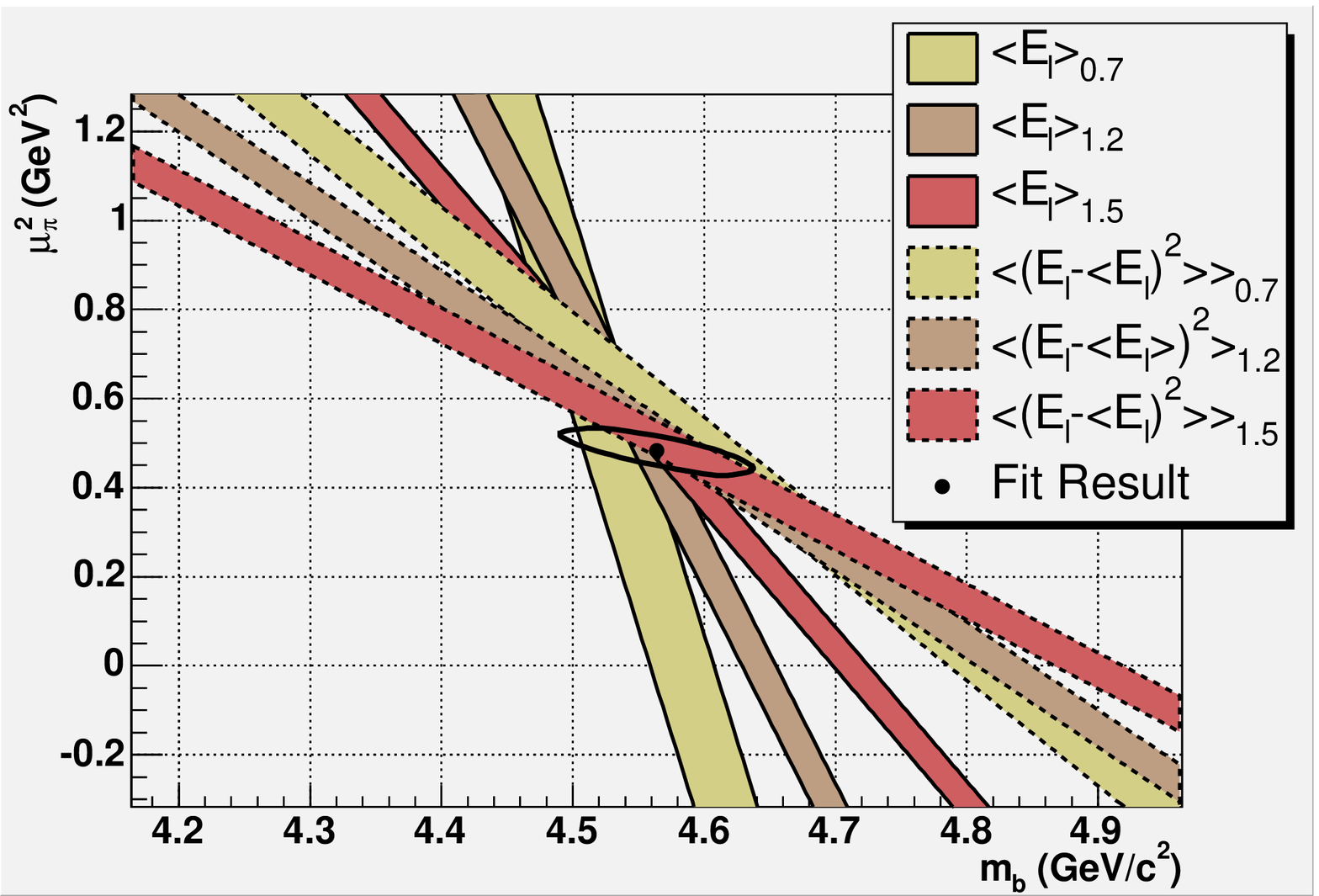,width=0.495\textwidth} \hfill
\epsfig{file=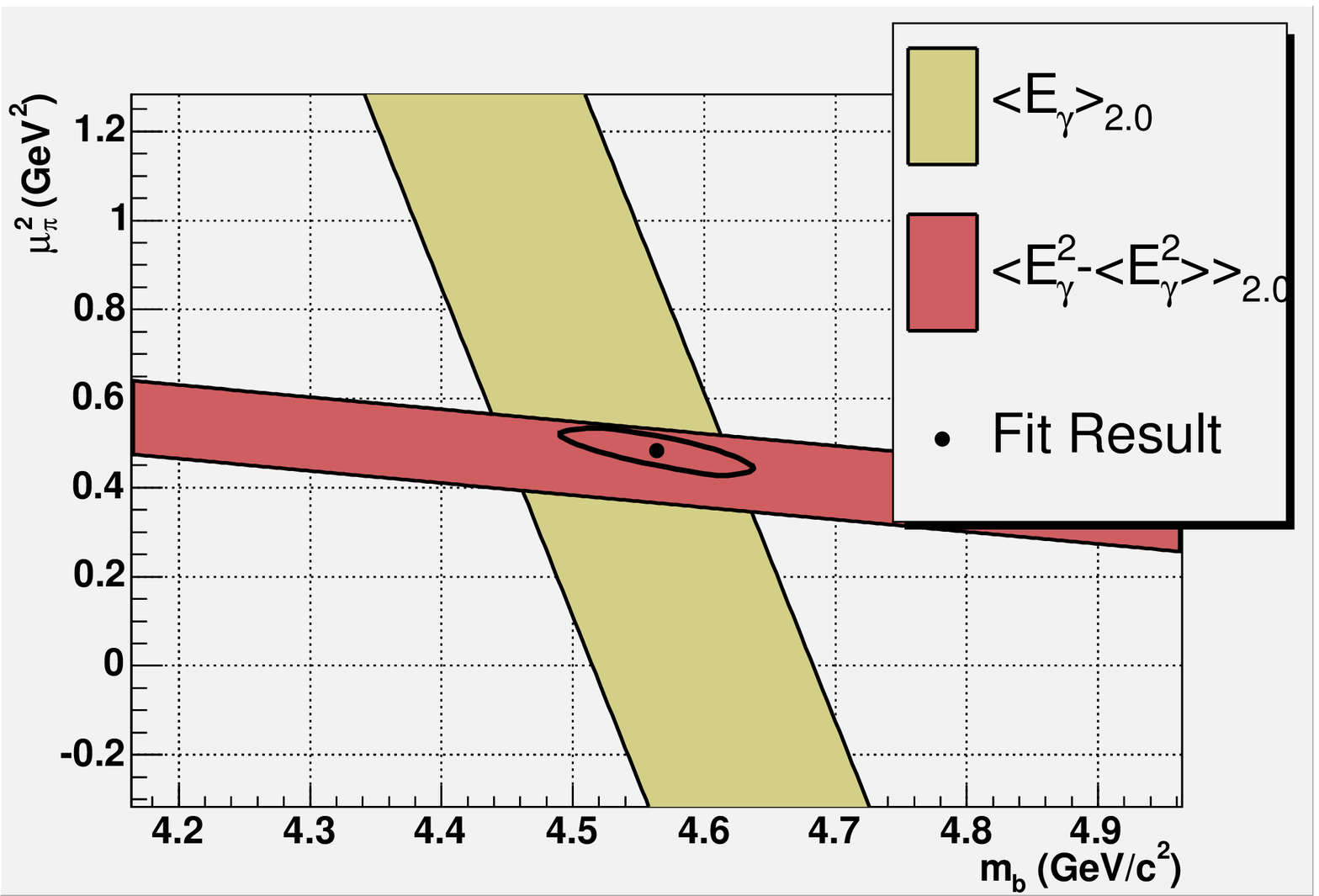,width=0.495\textwidth}
\caption{Constraint contours from various CLEO moment measurements on HQE
parameters $m_b$ and $\mu_\pi^2$.  In all plots, the ellipse shows the
$\Delta\chi^2=1$ contour from the fit to all moments in Table~\ref{tab:hqe}.
\label{fig:hqeConstraints}}
\end{figure}

The spectral moment measurements described in Sections \ref{sec:MX}
and \ref{sec:El}, and CLEO measurements of the $B\to X_s\gamma$ photon
energy spectrum~\cite{Eg} may be combined to make a global fit for the
Heavy Quark Expansion parameters, \cite{BCH04} following closely 
the theoretical treatment of Uraltsev and Gambino.\cite{HQE}
The combined fit to all moments given in Table~\ref{tab:hqe} has
reasonable confidence level, with $\chi^2=6.28/8$ d.o.f.  Each of the
moment measurements constrains the HQE parameters; we plot the allowed
regions for each measurement in Fig.~\ref{fig:hqeConstraints}.  The
consistency validates the use of the HQE to compute hadronic
corrections and measure $|V_{cb}|$ from inclusive measurements.
Preliminary results for the HQE parameters are
given in Table~\ref{tab:hqe}, which when combined with the branching
fraction result of Sec.~\ref{sec:El}, subtracting a small contribution
($\approx 1$\%) from $B\to X_u\ell\nu$, and the world average $B$
lifetime~\cite{PDG} yield $|V_{cb}|= 0.0424 \pm 0.0008$.  No
theoretical uncertainty is included in this preliminary result, but it
is expected to be comparable to the experimental uncertainty.
We note that the deviation of $\mu_G^2$ from the value expected from
the $B$--$B^*$ mass splitting ($0.35\pm0.07$ GeV$^2$) has negligible
effect on $|V_{cb}|$.

\section*{Acknowledgments}
The author wishes to thank the National Science Foundation for
research and travel support.

\end{document}